\documentclass[aps,prl,twocolumn,final,floatfix,superscriptaddress]{revtex4}
\usepackage{graphicx}
\usepackage{amssymb}
\usepackage{amsmath}

\newcommand{\nblock}{N_{\rm b}}
\newcommand{\rblock}{r_{\rm b}}
\newcommand{\tpulse}{t_{\rm exc}}

\begin{document}

\title{Strongly Coupled Plasmas via Rydberg-Blockade of Cold Atoms}

\author{G. Bannasch}
\affiliation{Max Planck Institute for the Physics of Complex Systems, D-01187 Dresden, Germany}
\author{T.C. Killian}
\affiliation{Rice University, Department of Physics and Astronomy and Rice Quantum
  Institute, Houston, Texas, 77251, USA}
\author{T. Pohl}
\affiliation{Max Planck Institute for the Physics of Complex Systems, D-01187 Dresden, Germany}

\begin{abstract}
We propose and analyze a new scheme to produce ultracold neutral plasmas deep in the strongly coupled regime. The method exploits the interaction blockade between cold atoms excited to high-lying Rydberg states and therefore does not require substantial extensions of current ultracold plasma experiments. Extensive simulations reveal a universal behavior of the resulting Coulomb coupling parameter, providing a direct connection between the physics of strongly correlated Rydberg gases and ultracold plasmas. The approach is shown to reduce currently accessible temperatures by more than an order of magnitude, which opens up a new regime for ultracold plasma research and cold ion-beam applications with readily available experimental techniques.
\end{abstract}

\maketitle
Strongly coupled Coulomb systems occupy an exotic regime of plasma physics \cite{ich82,red97,dun99} where correlated dynamics due to strong interactions dominates random thermal motion of the charges. 
While plasmas generally constitute the most abundant state of matter \cite{bat96},
strongly coupled systems are rather scarce and typically require extremely high densities
that occur in such exotic settings as the interior of stars and giant gas planets
\cite{vho91} or inertial-confinement fusion experiments \cite{amt04}. On the other hand,
ultracold neutral plasmas (UNPs), produced by direct photoionization of laser-cooled atoms
\cite{{kkb99,kpp07,kil07}}, represent a promising alternative for creating and studying
strongly coupled plasmas, which provides unique opportunities to probe dynamical phenomena
\cite{kpp07}. Their low temperatures yield strong coupling conditions at such small
densities that typical evolution timescales are slowed down tremendously. This has enabled
recent experiments to investigate collective excitations \cite{fzr06,cmk10}, relaxation
processes \cite{ber08,bcm12} as well as dynamical instabilities \cite{zfr08} at ultralow
plasma temperatures. Such conditions also hold great promise for applications in
nanotechnology, where UNPs are utilized to create high-brightness ion beams
\cite{gee07,han08,rei09,rei10,knu11,deb12} for fabricating and characterizing nanoscale
objects.

Yet, these intriguing possibilities are currently limited by intrinsic ion heating \cite{mur01,ger03,scg04,csl04,ppr05,cdd05,bdl11,lbm12} that takes place right after plasma creation. UNPs are produced from ultracold atoms whose initial temperatures suggest extremely strong correlations, which originally raised high hopes for creating neutral plasmas with unprecedented coupling strengths \cite{kkb99}. However, this scheme generates the plasma far from equilibrium in an entirely disordered state, and the subsequent correlation build-up is accompanied by substantial heating that pushes the plasma-ions just to the edge of the strong-coupling regime. Theoretical work has suggested ways to suppress this disorder-induced heating (DIH) by introducing initial ion correlations, e.g., by pre-ordering atoms in an optical lattice \cite{ppr04a} and by exploiting atomic correlations in a degenerate Fermi gas \cite{mur01} or correlations due to Penning ionization of excited atoms \cite{gem04}. These ideas set challenging experimental demands such as perfectly filled lattices with $\sim10^5$ atoms or very high atomic densities, which, thus far, have precluded their experimental implementation. 

\begin{figure}[t!]
 \includegraphics[width=1.0\columnwidth]{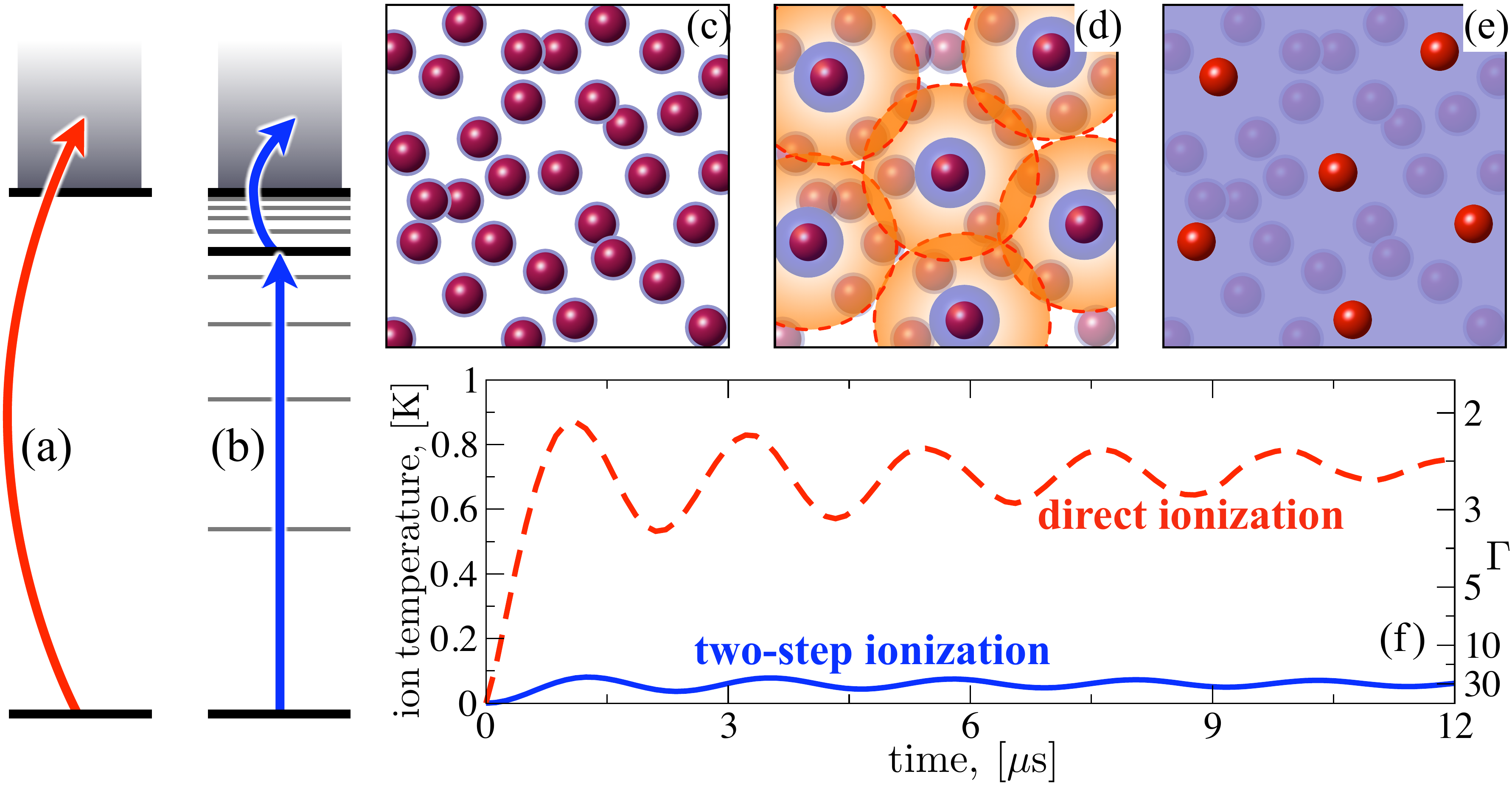}
 \caption{(Color online) Schematics of the (a) conventional direct-ionization scheme for UNP production and (b) the two-step production proposed in this work. Starting from a random gas of ground state atoms (c), the latter creates a correlated arrangement of Rydberg atoms (d), whose subsequent ionization results in a correlated plasma (e). As shown in panel (f), these initial correlations substantially suppress DIH after plasma creation (solid line) compared to the conventional production scheme (dashed line). The depicted temperature evolution has been calculated for an initial plasma density of $\rho=2\times10^8$cm$^{-3}$ and resonant excitation to Sr$(100^{1\!}S_{0})$ Rydberg states (see text for laser parameters).
 }
 \label{fig1}
\end{figure}

Here we describe a new UNP-production scheme that strongly suppresses DIH with currently available and well established experimental techniques. The approach is based on the strong correlations between Rydberg atoms that emerge when a cold gas of atoms is laser-excited to a high-lying electronic state. The frequency of the excitation laser turns out to permit continuous variation of the plasma coupling strength. For the first time, this will allow plasma temperatures and densities to be tuned independently, enabling unprecedented studies of the phase diagram \cite{sha11} of UNPs. Our calculations reveal a universal relation between the plasma's coupling strength and the Rydberg excitation dynamics, suggesting that UNPs could provide a powerful probe of the correlation properties of ultracold Rydberg gases. We demonstrate a maximum suppression of DIH by more than an order of magnitude, which opens up a new regime for UNP research and significantly boosts the brightness and resolution of UNP-based ion beam technology \cite{gee07}.

\begin{figure}[t]
 \includegraphics[width=1.0\columnwidth]{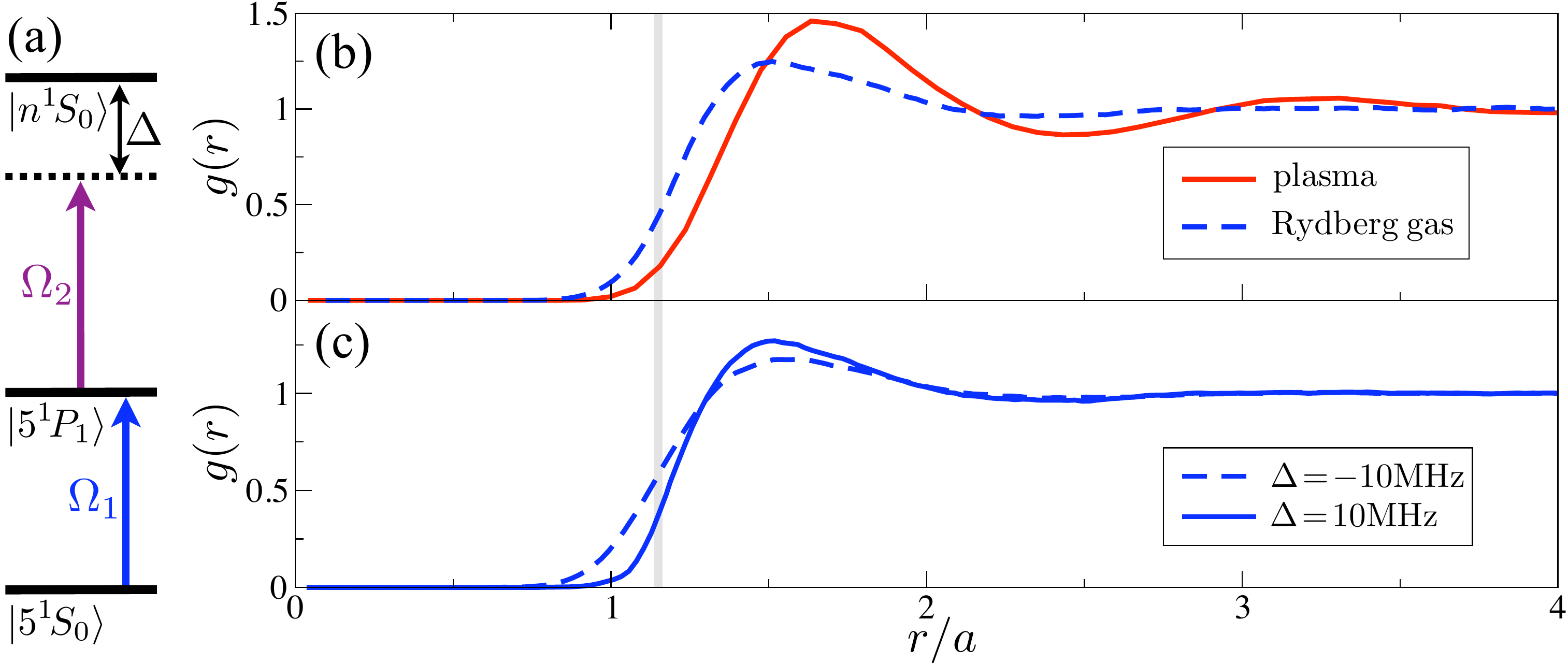}
 \caption{(Color online) (a) Schematic level diagram for two-photon Rydberg excitation. (b) Correlation function $g(r)$ for the Rydberg ensemble and the resulting plasma after relaxation for the same parameters as in Fig.\ref{fig1}(f). The vertical line marks the blockade radius according to eq.(\ref{eq:rblock}). (c) Correlation function of the Rydberg ensemble for two different laser detunings $\Delta$.
 }
 \label{fig2}
\end{figure}

Figure \ref{fig1} illustrates the basic idea in comparison to the conventional approach for UNP creation. 
In the latter case, direct photoionization of cold atoms [Fig.\ref{fig1}(a)] produces a disordered configuration of ions, whose subsequent rearrangement establishes inter-particle correlations, and thereby decreases the potential energy of the plasma. Since the total energy is conserved, this leads to a rapid rise of the temperature to $T_{\rm i}\sim1$\,K, as shown in Fig.\ref{fig1}(f). The corresponding degree of equilibrium correlations can be characterized by the so-called Coulomb coupling parameter $\Gamma = \frac{e^2}{a \, k_\text{B} T}$, where $e$ is the electron charge, $k_{\rm B}$ is the Boltzmann constant, and $a = (\frac{4}{3}\pi\rho)^{-\frac{1}{3}}$ is the Wigner-Seitz radius for a plasma of density $\rho$.  A plasma is termed strongly coupled, when the average potential energy $\sim e^2/a$ of the charges exceeds their thermal energy $\sim k_{\rm B}T$, i.e. when $\Gamma > 1$. Starting from a random distribution of almost stationary ions, the subsequent equilibration and DIH establishes a Coulomb coupling parameter of $\Gamma\approx2$ irrespective of the initial ion density and atom temperature \cite{csl04,ppr05,lbm12} [cf. Fig.\ref{fig1}(f)]. This strong heating can be reduced with a two-step ionization scheme [see Fig.\ref{fig1}(b)], where the ground state atoms are first excited to high-lying Rydberg states and subsequently ionized. In the first excitation step we take advantage of the so called Rydberg blockade \cite{lfc01}, which prevents simultaneous excitation of nearby atoms due to the enormous van der Waals interaction between Rydberg atoms \cite{gal94}. As illustrated in Fig.\ref{fig1}(d), this gives rise to strong Rydberg-Rydberg atom correlations (see also Fig.\ref{fig2}). Their subsequent ionization thus produces a pre-correlated plasma [Fig.\ref{fig1}(e)] that evolves to an equilibrium state with a much higher Coulomb coupling parameter [Fig.\ref{fig1}(f)]. Indeed the final plasma correlation function, shown in Fig.\ref{fig2} closely resembles that of the prepared Rydberg gas but also shows pronounced oscillations characteristic of a strongly coupled Coulomb liquid.

The Rydberg blockade effect has been the basis of recent experimental breakthroughs for applications in quantum information \cite{urb09,gae09,swm10} and nonlinear quantum optics \cite{pmg10,duk12,pey12,max12,par12,hof12}, and opened the door for  studies of quantum many-body phenomena \cite{rai08,sch10,vit11,sch11,sch12,vit12}. While most of these experiments utilize alkaline atoms, we focus here on cold Strontium atoms whose secondary valence electron provides a viable way to probe the temperature of the created plasma ions via optical absorption and fluorescence measurements \cite{scg04,cdd05,lah07}. Following recent experiments on Sr-Rydberg gases \cite{mil10,mcq13,loc12}, we consider resonant two-photon excitation from the Sr$(5^{1\!}S_0)$ ground state to a Sr$(n^{1\!}S_0)$ Rydberg state via the intermediate Sr$(5^{1\!}P_1)$ state by two excitation lasers with Rabi frequencies $\Omega_1$ and $\Omega_2$ and a two-photon detuning $\Delta$ [see Fig.\ref{fig2}(a)]. The single-atom steady state excitation fraction $P_{\rm ryd}=\Omega_1^2/(\Omega_1^2+\Omega_2^2)$ needs to be high in order to maximize interaction effects. We, therefore, choose $\Omega_1=25$MHz and $\Omega_2=0.5$MHz, which gives $P_{\rm ryd}\approx1$ and can be realized in current experiments \cite{mil10,mcq13}. Due to the large decay rate $\gamma=32$MHz of the intermediate state, the resulting many-body excitation dynamics can be described in terms of transition rates \cite{app07,asp11}. The corresponding set of rate equations can be solved efficiently via classical Monte Carlo simulations that account for the Rydberg-Rydberg atom interaction via an effective detuning \cite{app07,asp11}
\begin{equation}
  \label{eq:detuning}
  \tilde{\Delta}_i = \Delta - \sum_{j\ne i} \frac{C_6}{|{\bf r}_i-{\bf r}_j|^6}
\end{equation}
of the $i$th atom at position ${\bf r}_i$, where the sum runs over all atoms that are in the Rydberg state. This approach has been applied to study cold Rydberg gases \cite{app07,asp11,hge12} and lattices \cite{app07,hge12,hoe12,igo12} and was shown to yield good agreement with quantum calculations \cite{asp11,hoe12}. For large principal quantum numbers, $n$, the van der Waals coefficient $C_6$ of Sr$(n^{1\!}S_0)$ Rydberg states \cite{mmn11,vjp12} takes on enormous values such that a single Rydberg atom can inhibit the excitation of up to a few $10^3$ atoms. This implies a large number of ground state atoms required to obtain sufficiently many Rydberg excitations. Such sizable ensembles can be described by the Monte Carlo procedure, which allows us to treat up to several $10^7$ atoms. Typical simulation results are illustrated in Fig.\ref{fig2}, where we show the Rydberg-Rydberg atom correlation function after laser excitation. At short distances, simultaneous excitation is completely suppressed by the interaction blockade \cite{lfc01}. The corresponding blockade radius 
\begin{equation}
  \label{eq:rblock}
\rblock=  \left(\frac{C_6}{\hbar\Delta \nu}\right)^{1/6}
\end{equation}
can be estimated by equating the respective interaction energy with the linewidth $\Delta\nu$ \footnote{For the parameters of this work ($\Omega_1=25$MHz, $\Omega_2=0.5$MHz, $\tpulse=0.5\mu$s) $\Delta\nu=13.7$MHz.} of the Rydberg excitation. This simple relation yields a good description of the numerical results in Fig.\ref{fig2} and generally gives $\rblock \approx 1.2 \, a$, which is close to the typical correlation length of a strongly coupled plasma.

\begin{figure}[t]
 \includegraphics[width=1.0\columnwidth]{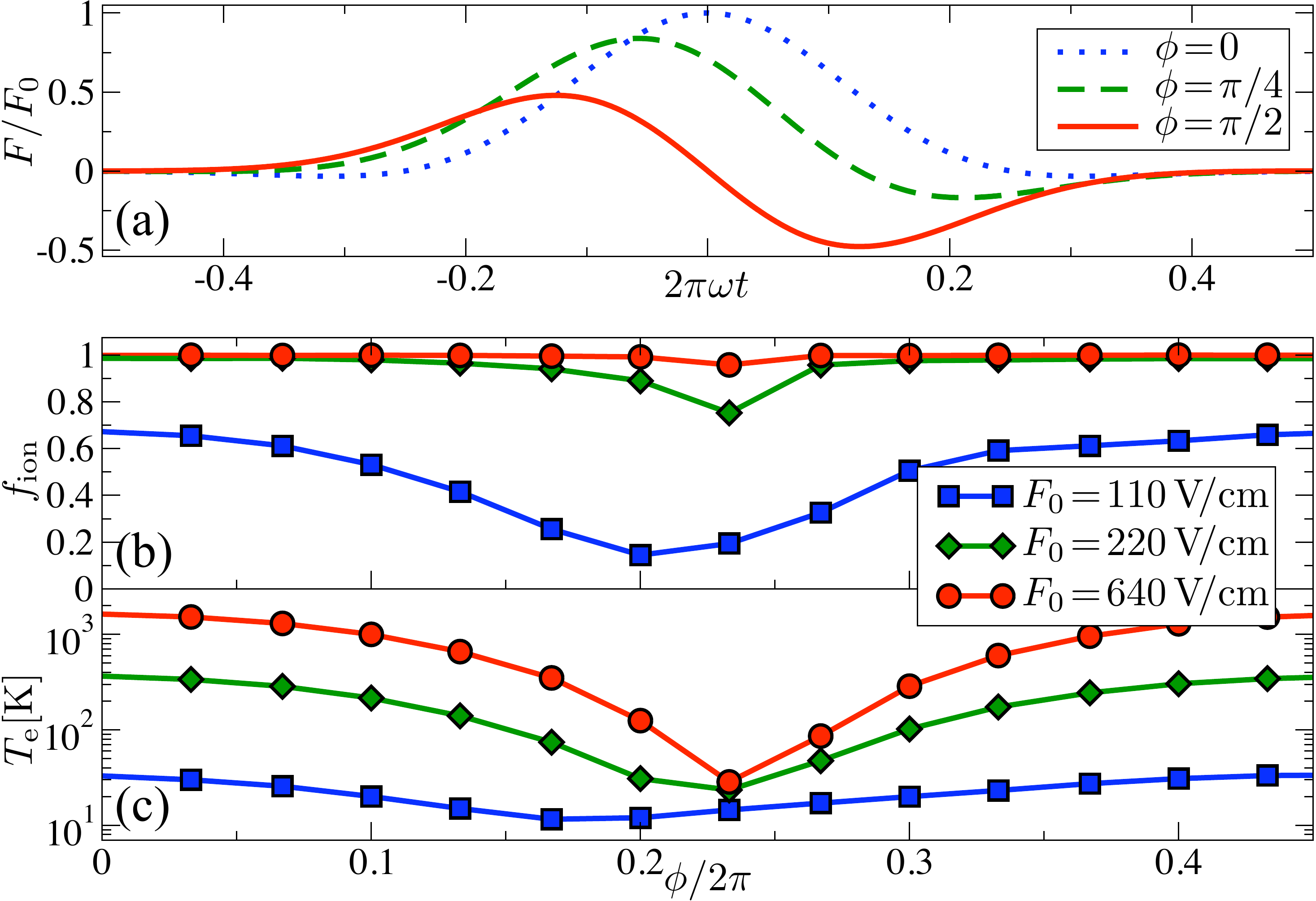}
 \caption{(Color online) (a) Considered shape of the field pulse for Rydberg atom ionization with $\tau=2\pi/5\omega$.
   (b) Ionization fraction, $f_{\rm ion}$, and (c) electron excess energy, $E_{\rm e}$, for $(2\pi\omega)^{-1} = 100$\,ps, $n = 70$ and different values of $F_0$. 
 }
 \label{fig3}
\end{figure}

The conversion of these atomic correlations into ionic ones, requires efficient ionization of the Rydberg states while leaving the ground state atoms undisturbed. To this end, we consider ionization by a short electric field pulse, which we describe by classical trajectory Monte-Carlo (CTMC) simulations \cite{per79}. We assume a simple pulse shape $F=F_0\cos(\omega \, t + \phi) \exp(- t^2/\tau^2)$, where $F_0$ denotes the field amplitude and $\omega$, $\phi$ and $\tau$ parametrize the form of the pulse [cf. Fig.\ref{fig3}(a)]. The ionization processes has to fulfill two major requirements: (i) ionization needs to be efficient to maximize the transfer of the Rydberg atom ordering into ionic correlations and (ii) the excess energy $E_{\rm e}$ of the ionized electrons should be sufficiently low so that they remain trapped in the collective ionic space charge potential to form a neutral plasma state \cite{kkb99}. As shown in Fig.\ref{fig3}(b) both conditions are optimized around $\phi\approx\pi/2$, where we find a low electron energy $\sim 100$K and nearly $100\%$ ionization of Sr($70^{1\!}S_0$) Rydberg states. This energy is low enough to enable plasma formation \cite{kkb99} but sufficiently high to prevent rapid recombination after plasma creation \cite{klk01,fzr07,gls07,pvs08,bap11}. Such ideal conditions can also be obtained with other pulse shapes, such as, e.g., half-cycle pulses as shown theoretically and experimentally in \cite{wgn02}. 

Finally, we performed molecular dynamics simulations \cite{fmm} to investigate the subsequent 
relaxation of the plasma, which is described as a one-component system \footnote{For our typical electron temperatures and densities, electronic screening does not affect the ion temperature evolution \cite{csl04} and can, therefore, be neglected.} of initially stationary ions, whose initial positions correspond to those of the Rydberg atoms obtained from the Monte-Carlo simulations of the excitation step. Such  simulations yield the temperature dynamics shown in Fig.\ref{fig1}(f), which demonstrates a more than tenfold enhancement of the Coulomb coupling parameter. This places the produced plasma deep into the strongly coupled regime with strong liquid like correlations [see Fig.\ref{fig2}(a)].

\begin{figure}[b!]
 \includegraphics[width=1.0\columnwidth]{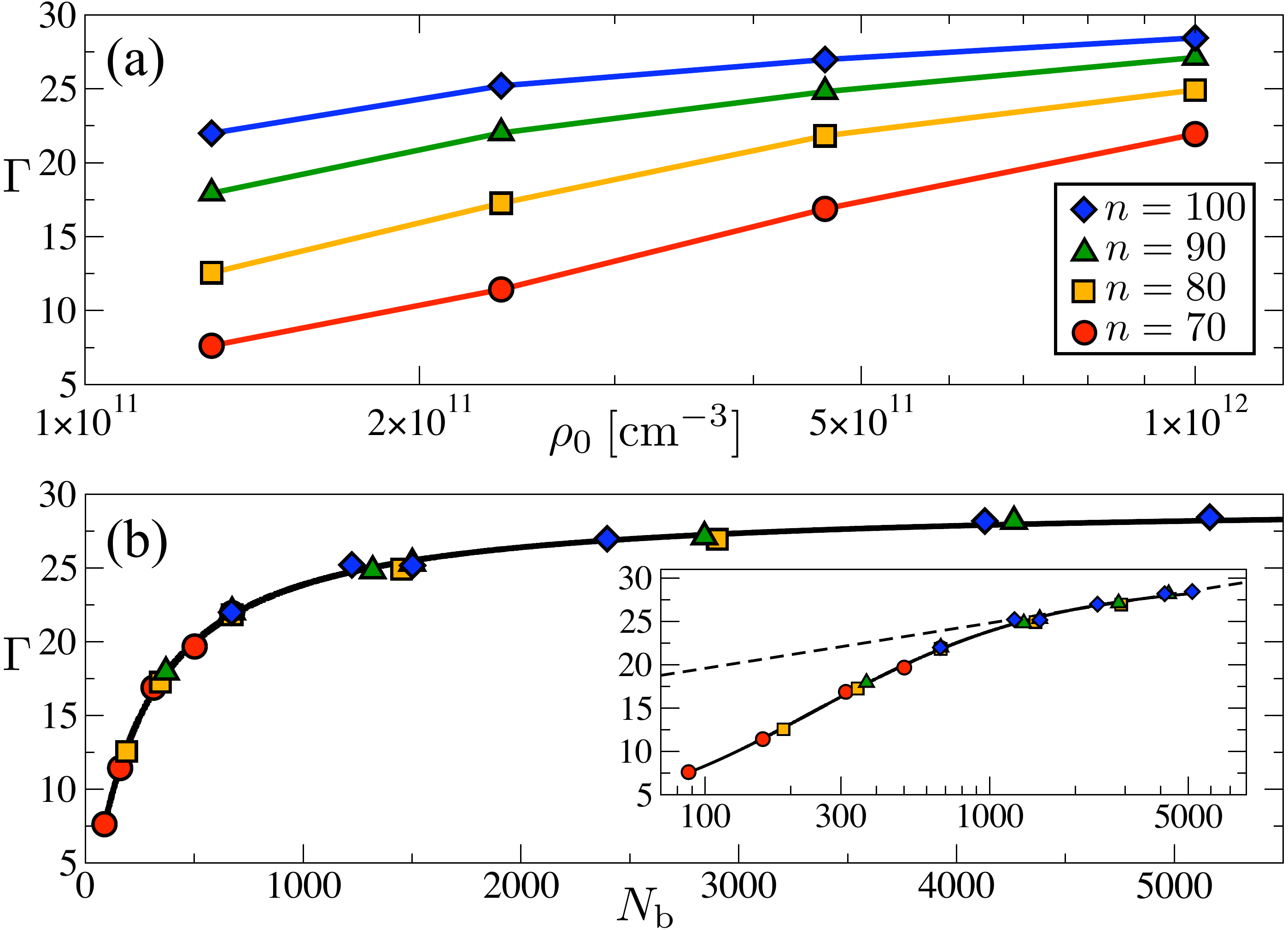}
 \caption{(Color online) (a) Final Coulomb coupling parameter $\Gamma$ as a function of the ground state atom density $\rho_0$ for $\tpulse = 0.5 \, \mu$s and different Rydberg levels $n$. (b) Same data, but as a function of the number $\nblock$ [eq.(\ref{eq:nblock})] of blocked atoms. The inset demonstrates the logarithmic scaling (dashed line) of $\Gamma$.
 }
 \label{fig4}
\end{figure}

The described production scheme involves several parameters that can be widely tuned experimentally. To elucidate their effects we have determined the achievable coupling strengths $\Gamma$ for various combinations of the ground state atom density $\rho_0$, the addressed Rydberg level $n$ and the detuning $\Delta$ of the Rydberg excitation laser.
For fixed $\tpulse=0.5\mu$s and resonant excitation, the Coulomb coupling parameter $\Gamma$ increases with both the atomic density $\rho_0$ and the principal quantum number $n$ of Rydberg state [Fig.\ref{fig4}(a)]. This behavior can be qualitatively understood from the increasing degree of correlations in cold Rydberg gases with $\rho_0$ and the strength of the van der Waals interactions. We can quantify this relation by defining the number of blockaded atoms 
\begin{equation}
  \label{eq:nblock}
  \nblock = \frac{4}{3} \pi \rblock^3 \, \rho_0\;,
\end{equation}
which corresponds to the total number of atoms within one blockade sphere of radius $\rblock$. Indeed all data points collapse on a single universal curve as a function of $\nblock$ [see Fig.\ref{fig4}(b)]. 
In the strong blockade limit $\Gamma$ shows a weak logarithmic growth with $\nblock$ such that further increase of the ground state atom density does not yield a substantial enhancement.

\begin{figure}[t]
 \includegraphics[width=1.0\columnwidth]{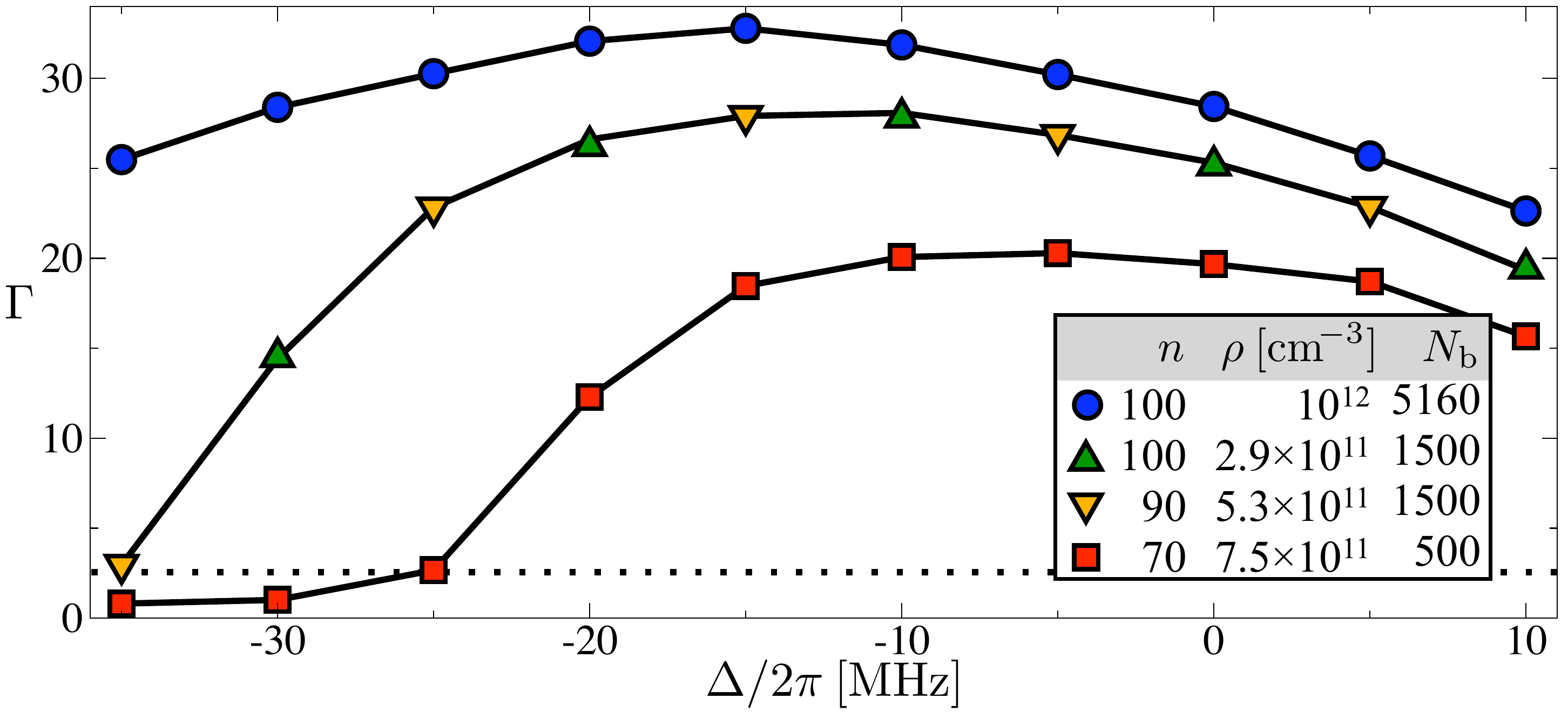}
 \caption{(Color online) Coulomb coupling parameter $\Gamma$ as a function of laser detuning $\Delta$ for $\tpulse = 0.5 \, \mu$s and different combinations of $n$ and $\rho_0$. The horizontal dotted line shows the Coulomb coupling parameter for direct ionization.}
 \label{fig5}
\end{figure}

However, the detuning of the Rydberg excitation laser provides another control parameter to tune the Coulomb coupling strength. This is demonstrated in Fig.\ref{fig5} where we show $\Gamma$ as a function of $\Delta$ for different ground state densities and Rydberg states. Apparently the universal scaling also holds off resonance, such that the final result is determined by $\Delta$ and $\nblock$ only.
For the attractive interactions between Sr($n^{1\!}S_0$) atoms \cite{mmn11,vjp12} a finite laser detuning gives rise to interaction-induced resonances at the red side of the excitation line ($\Delta<0$), where Rydberg excitation is enhanced for configurations where $\tilde{\Delta}_i\approx0$. This additional distance-selectivity can produce a more pronounced peak structure of the correlation function \cite{rob05} and tends to steepen its drop around $\rblock$ [see Fig.\ref{fig2}(b)]. As a result the Coulomb coupling parameter can be increased by tuning to the red side of the atomic resonance. For large $\nblock$ it assumes a maximum around $\Delta=-\gamma/2$ and drops at either side of the atomic line. Well off resonance, DIH is even enhanced, such that $\Gamma$ can be tuned from the weak-to-moderate coupling region deep into the strong coupling regime with maximal Coulomb coupling parameters around $\Gamma\sim35$.

The proposed scheme, therefore, represents a viable way to not only tremendously boost
currently achievable coupling strength but also to independently tune the plasma
parameters of UNPs, which would otherwise be constrained to $\Gamma=2\propto \rho^{1/3}T$.
The demonstrated tenfold cooling directly translates into an enhanced brightness of
UNP-based ion beams, which enhances the spatial beam resolution \cite{gee07} for
nanotechnology applications. 

We have focussed on electric-field ionization of the Rydberg gas, but other approaches for plasma-conversion should also be possible, such as long-wavelength photoionization or electron impact ionization. As pointed out recently \cite{rod12}, the latter occurs naturally in dense Rydberg gases, driving a spontaneous avalanche-like evolution to a plasma \cite{rob00,rob03,ppr03,mrk08,web12} that could inherit residual Rydberg atom correlations. Spontaneous plasma formation, however, takes place on long microsecond timescales \cite{rob00,rod12} and involves significant atomic motion \cite{arw07} and loss of highly-excited atoms to low-lying states \cite{rob03,ppr03} $-$ all of which deteriorate the cooling effect described  here. However, electron impact ionization can be driven with less deleterious effect by combining Rydberg excitation with the conventional method for plasma creation \cite{kkb99} in a double-pulse scheme. Here, the first laser-pulse is tuned below threshold to excite strongly correlated Rydberg atoms while leaving the majority of  atoms in the ground state. The second pulse, tuned above threshold, ionizes a fraction of the ground state atoms to produce a low-density plasma with energetic electrons that subsequently convert the Rydberg gas into a correlated plasma. Such double-pulse sequences have already been implemented for pump-probe experiments in Rydberg gases \cite{rei08,sch10}, to excite Rydberg atoms in a UNP \cite{van05,pcz06}, and to convert a Rydberg gas to a plasma \cite{mcq13}.  The final degree of plasma correlations will depend on the parameters of the seed-plasma, which raises interesting questions for future studies.

Several extensions of the proposed scheme appear worth pursuing. First, a correlated plasma with $T\lesssim100$mK seems well suited for subsequent laser-cooling on the ${\rm Sr}^{\!+}\!(5S_{1/2})\!\leftrightarrow\!{\rm Sr}^{\!+}\!(5P_{1/2})$ transition, which thus far was hampered by DIH out of the Doppler range. Second, long-range dipolar Rydberg interactions, induced by resonant pair-state couplings \cite{vog06,rcy08,yrp09} or external electric fields \cite{vog07}, may give rise to atomic correlations that are closer to those of long-range interacting Coulomb systems and, therefore, may further enhance the coupling strengths of the resulting plasma. Here, higher order interactions give rise to an interesting, correlated excitation dynamics \cite{yrp09,pob09,dan12} that could be probed by monitoring $\Gamma$ of the corresponding UNP. Finally, Rydberg excitation by standing-wave light fields is expected to increase atomic correlations substantially. If the lattice constant matches the blockade radius, we expect strong excitation ordering that could push $\Gamma$ to values not far below the crystalline-plasma regime ($\Gamma\approx174$ \cite{ich82}). 

This work was supported by the Department of Energy Partnership in Basic Plasma Science
and Engineering (PHY-1102516) and the Air Force Office of Scientific Research
(FA9550-12-1-0267), the EU through the Marie Curie ITN "COHERENCE".

\end{document}